\documentclass[twocolumn,tighten,times]{aastex61}
\usepackage{upgreek}
\usepackage{newtxtext}
\usepackage{newtxmath}
\usepackage{bm,url}

\begin{document}

\received{\today}
\revised{\today}
\accepted{\today}
\submitjournal{ApJL}

\title{LOFAR discovery of the fastest-spinning millisecond pulsar in the Galactic field}
\shorttitle{Fastest-spinning millisecond pulsar in the Galactic field}
\shortauthors{Bassa et al.}

\correspondingauthor{C.~G.~Bassa}
\email{bassa@astron.nl}

\author[0000-0002-1429-9010]{C.~G.~Bassa}
\affiliation{ASTRON, the Netherlands Institute for Radio Astronomy, Postbus 2, NL-7990 AA Dwingeloo, The Netherlands}

\author{Z.~Pleunis}
\affiliation{Department of Physics and McGill Space Institute, McGill University, 3600 University St., Montreal, QC H3A 2T8, Canada}

\author{J.~W.~T.~Hessels}
\affiliation{ASTRON, the Netherlands Institute for Radio Astronomy, Postbus 2, NL-7990 AA Dwingeloo, The Netherlands}
\affiliation{Anton Pannekoek Institute for Astronomy, University of Amsterdam, Science Park 904, 1098 XH Amsterdam, The Netherlands}

\author{E.~C.~Ferrara}
\affiliation{Center for Research and Exploration in Space Science, NASA Goddard Space Flight Center, Greenbelt, MD 20771, USA}
\affiliation{Department of Astronomy, University of Maryland, College Park, MD 20742, USA}

\author{R.~P.~Breton}
\affiliation{Jodrell Bank Centre for Astrophysics, School of Physics and Astronomy, The University of Manchester, Manchester M13\,9PL, UK}

\author{N.~V.~Gusinskaia}
\affiliation{Anton Pannekoek Institute for Astronomy, University of Amsterdam, Science Park 904, 1098 XH Amsterdam, The Netherlands}

\author{V.~I.~Kondratiev}
\affiliation{ASTRON, the Netherlands Institute for Radio Astronomy, Postbus 2, NL-7990 AA Dwingeloo, The Netherlands}
\affiliation{Astro Space Centre, Lebedev Physical Institute, Russian Academy of Sciences, Profsoyuznaya Str.\ 84/32, Moscow 117997, Russia}

\author{S.~Sanidas}
\affiliation{Anton Pannekoek Institute for Astronomy, University of Amsterdam, Science Park 904, 1098 XH Amsterdam, The Netherlands}

\author{L.~Nieder}
\affiliation{Albert-Einstein-Institut, Max-Planck-Institut f\"ur Gravitationsphysik, D-30167 Hannover, Germany}
\affiliation{Leibniz Universit\"at Hannover, D-30167 Hannover, Germany}

\author{C.~J.~Clark}
\affiliation{Albert-Einstein-Institut, Max-Planck-Institut f\"ur Gravitationsphysik, D-30167 Hannover, Germany}
\affiliation{Leibniz Universit\"at Hannover, D-30167 Hannover, Germany}

\author{T.~Li}
\affiliation{Key Laboratory of Optical Astronomy, National Astronomical Observatories Chinese Academy of Sciences, 100012, Beijing, China}
\affiliation{Isaac Newton Group of Telescopes, Apartado de correos 321, Santa Cruz de La Palma, E-38700, Spain}

\author{A.~S.~van~Amesfoort}
\affiliation{ASTRON, the Netherlands Institute for Radio Astronomy, Postbus 2, NL-7990 AA Dwingeloo, The Netherlands}

\author{T.~H.~Burnett}
\affiliation{Department of Physics, University of Washington, Seattle, WA 98195-1560, USA}

\author{F.~Camilo}
\affiliation{Square Kilometre Array South Africa, Pinelands, 7405, South Africa}

\author{P.~F.~Michelson}
\affiliation{W. W. Hansen Experimental Physics Laboratory, Kavli Institute for Particle Astrophysics and Cosmology, Department of Physics and SLAC National Accelerator Laboratory, Stanford University, Stanford, CA 94305, USA}

\author{S.~M.~Ransom}
\affiliation{National Radio Astronomy Observatory, 1003 Lopezville Road, Socorro, NM 87801, USA}

\author{P.~S.~Ray}
\affiliation{Space Science Division, Naval Research Laboratory, Washington, DC 20375-5352, USA}

\author{K.~Wood}
\affiliation{Praxis Inc., Alexandria, VA 22303, resident at Naval Research Laboratory, Washington, DC 20375, USA}

\begin{abstract}
  We report the discovery of PSR\,J0952$-$0607, a 707-Hz binary
  millisecond pulsar which is now the fastest-spinning neutron star
  known in the Galactic field (i.e., outside of a globular cluster).
  PSR\,J0952$-$0607 was found using LOFAR at a central observing
  frequency of 135\,MHz, well below the 300\,MHz to 3\,GHz frequencies
  typically used in pulsar searches.  The discovery is part of an
  ongoing LOFAR survey targeting unassociated \textit{Fermi} Large
  Area Telescope $\gamma$-ray sources.  PSR\,J0952$-$0607 is in a
  6.42-hr orbit around a very low-mass companion
  ($M_\mathrm{c}\gtrsim0.02$\,M$_\odot$) and we identify a strongly
  variable optical source, modulated at the orbital period of the
  pulsar, as the binary companion. The light curve of the companion
  varies by 1.6\,mag from $r^\prime=22.2$ at maximum to
  $r^\prime>23.8$, indicating that it is irradiated by the pulsar
  wind.  {\it Swift} observations place a 3-$\sigma$ upper limit on
  the $0.3-10$\,keV X-ray luminosity of $L_X < 1.1 \times
  10^{31}$\,erg\,s$^{-1}$ (using the 0.97\,kpc distance inferred from
  the dispersion measure).  Though no eclipses of the radio pulsar are
  observed, the properties of the system classify it as a black widow
  binary.  The radio pulsed spectrum of PSR\,J0952$-$0607, as
  determined through flux density measurements at 150 and 350\,MHz, is
  extremely steep with $\alpha\sim-3$ (where $S \propto
  \nu^{\alpha}$). We discuss the growing evidence that the
  fastest-spinning radio pulsars have exceptionally steep radio
  spectra, as well as the prospects for finding more sources like
  PSR\,J0952$-$0607.
\end{abstract}


\section{Introduction}
The discovery of the first millisecond pulsar (MSP), PSR\,B1937+21
with a spin frequency of 642\,Hz, by \citet{bkh+82} came as a great
surprise, and demonstrated that some neutron stars can reach
astounding rotational rates.  The more recent discovery of {\it
  transitional} millisecond pulsars (tMSPs), which transition back and
forth between a rotation-powered MSP and an accretion-powered low-mass
X-ray binary (LMXB) state \citep{asr+09,pfb+13,bph+14}, confirmed the
basic recycling model of \citet{acrs82} and \citet{rs82} in which a
neutron star is spun-up to millisecond spin periods due to the
accretion of matter and angular momentum.  At the same time, the tMSPs
have also raised many questions about the detailed physics of the
pulsar recycling process and how efficient it can ultimately be in
terms of spinning-up neutron stars
(e.g.\ \citealt{dmm+15,abp+15,pmb+15,jah+16}).

It is striking that, since the discovery of PSR\,B1937+21, only one
faster-spinning MSP has been found (PSR\,J1748$-$2446ad, in the
globular cluster Terzan~5, spinning at 716\,Hz; \citealt{hrs+06}).
While the neutron star equation-of-state in principle allows spin
frequencies up to 1200\,Hz \citep{cst94,lp04} before mass-shedding or
break-up, the currently observed spin frequency distribution of radio
and X-ray MSPs cuts off around 730\,Hz
\citep{cmm+03,fw07,chak08,hess08}.  The question thus remains: can
nature form {\it sub}-millisecond pulsars?

Physical effects such as decoupling of the Roche lobe \citep{tau12},
transient accretion \citep{bc17}, and gravitational wave emission
\citep{cmm+03}, have been put forward to explain the observed spin
frequency cut-off.  Observationally, there are additional challenges
in detecting sub-millisecond pulsars, compared to canonical MSPs with
spin frequencies between 200 and 500\,Hz, but the cut-off around
730\,Hz is hard to explain purely as an observational bias.  For
example, computational advances allow present day radio pulsation
surveys to retain sensitivity to spin frequencies well in excess of
1000\,Hz (e.g.\ \citealt{lbh+15}) because it is now possible to record
data with sufficient time and frequency resolution -- which is
critical for correcting for the dispersive delays introduced by the
ionized interstellar medium (IISM).  Furthermore, if MSP searches are
conducted at sufficiently high radio frequencies (1 to 2\,GHz) the
effects of scattering in IISM should also not preclude the detection
of sub-millisecond pulsars, though high-frequency radio searches are
disadvantaged by the fact that pulsars typically have steep radio
spectra ($S \propto \nu^{\alpha}$, where $\alpha=-1.4\pm1.0$;
\citealt{blv13}).

\citet{wex+15} suggest that a possible bias against finding rapidly
spinning pulsars, and hence energetic MSPs, may be the irradiation
driven mass-loss from the binary companion, which can lead to eclipses
of the radio signal during large parts of the orbit
\citep{sah+14,rrb+15}.  Similar considerations were previously
presented by \citet{tava91}, motivated by the discovery of the first
eclipsing black widow pulsar binary system, PSR\,B1957+20, in which
the low-mass, bloated companion star is irradiated by the pulsar wind
\citep{fst88}.  Indeed, there is evidence that the eclipsing MSP
systems -- both the black widows with very-low-mass companions and the
redbacks with higher-mass ($M_\mathrm{c} \gtrsim 0.2$\,M$_\odot$)
companions -- are on average spinning faster than `classical' MSPs
with white dwarf companions \citep{hess08,ptrt14}.  For eclipsing
systems, there is again an advantage towards observing at higher radio
frequencies, where the eclipse durations are typically lower
\citep{asr+09}, but also the disadvantage that the intrinsic pulsar
spectrum is generally falling off rapidly towards higher frequencies.

Recent results by \citet{kvl+15}, \citet{kvh+16} and \citet{fjmi16}
indicate that the fastest-spinning MSPs tend to have the steepest
radio spectra ($\alpha < -2.5$), pointing to another possible bias
against finding fast-spinning MSPs in ongoing surveys, which focus on
central observing frequencies around 350\,MHz \citep{dsm+13,slr+14}
and 1.4\,GHz \citep{cfl+06,kjs+10,bck+13}. As a result, radio
pulsation searches at frequencies below 300\,MHz have the potential of
opening up a so far largely unexplored parameter space, in the cases
where IISM scattering is low and eclipsing does not hinder detection
either.

Here we present the discovery of PSR\,J0952$-$0607, a
very-steep-spectrum MSP, which is now the fastest-spinning neutron
star known in the Galactic field (outside of a globular cluster).
PSR\,J0952$-$0607 was found in a radio pulsation survey using the
Low-Frequency Array (LOFAR; \citealt{hwg+13,sha+11}) to target
unassociated \textit{Fermi} $\gamma$-ray sources. This {\it
  Fermi}-targeted approach has been successful in finding many new
MSPs \citep{rap+12}, but our survey is the first to use LOFAR to
survey at observing frequencies of 135\,MHz (see also
\citealt{pbh+17}). To enable this survey, a combination of coherent
and incoherent dedispersion has been employed to limit the effects of
dispersive smearing \citep{bph17}. In \S\,\ref{sec:observations} we
will highlight the discovery and multi-wavelength follow up, with the
results being presented in \S\,\ref{sec:results}. We discuss the
broader implications of PSR\,J0952$-$0607's discovery and conclude
this manuscript in \S\,\ref{sec:discussion}.

\begin{figure}
  \includegraphics[width=\columnwidth]{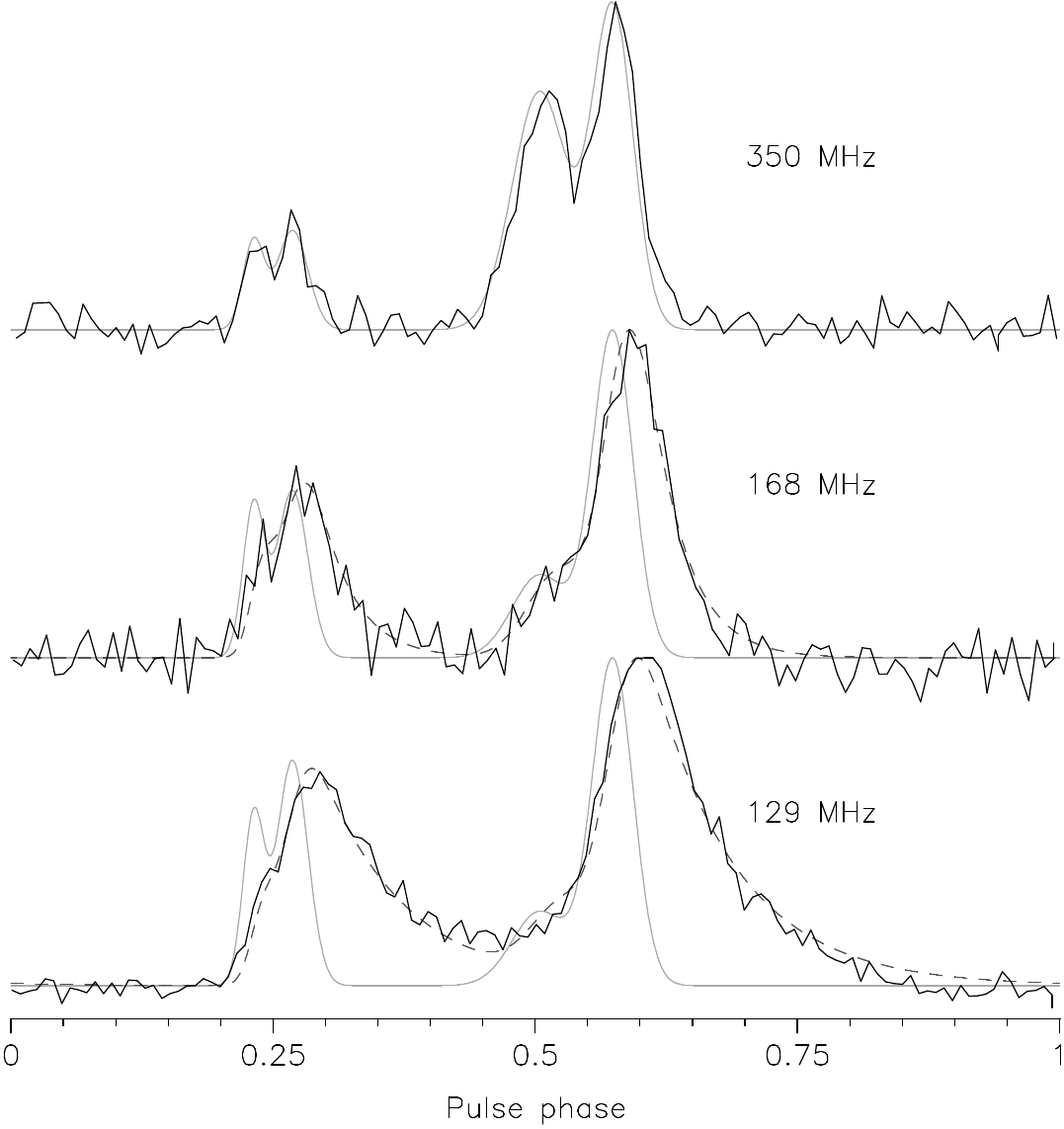}
  \caption{Integrated pulse profiles of PSR\,J0952$-$0607 at three
    observing frequencies (black; GBT: 350\,MHz, LOFAR: 168 and
    129\,MHz). The dashed grey lines are fits to the observed
    profiles. These assume thin screen scattering and an intrinsic
    profile consisting of von Mises functions fitted to the four
    components in the 350\,MHz profile. The positions and widths of
    the components are kept fixed, while the amplitudes are allowed to
    vary. The unscattered model profiles are shown with solid grey
    lines. All profiles are scaled to the same peak value for clarity.
    It is clear that the pulse profile becomes visibly scattered
    towards the bottom of the LOFAR HBA band. }
  \label{fig:profile}
\end{figure}

\section{Observations and analysis}\label{sec:observations}
\subsection{Radio}\label{ssec:radio}
PSR\,J0952$-$0607 was discovered as part of an ongoing LOFAR survey at
135\,MHz, continuing on the pilot survey by
\citet{pbh+17}. Unassociated $\gamma$-ray sources were selected from
an all-sky source list based on 7\,years of \textit{Fermi}-LAT (Large
Area Telescope; \citealt{aaa+09lat}) Pass\,8 data. That list resulted
from a preliminary version of the procedure that will be used to
produce the next public release LAT source catalog. Amongst these is a
new $\gamma$-ray source which has a test statistic of $~100$ and a
relatively small error radius of $3\farcm8$. The source's $\gamma$-ray
spectrum is strongly curved, peaking at 1.4\,GeV, and no significant
$\gamma$-ray emission is detected above 10\,GeV. As the source lies
well out of the plane ($b=35\fdg4$), these characteristics made it a
prime millisecond pulsar candidate.

The $\gamma$-ray source was observed for 20\,min on 2016 December 25
with LOFAR. The high-band antennas (HBAs) from the innermost 21 LOFAR
core stations (longest baseline of 2.3\,km) were used to form 7
tied-array beams ($3\farcm5$ FWHM), each covering a 39\,MHz wide band
centered at 135\,MHz. This setup is identical to that of
\citet{pbh+17}. The complex voltage output for each tied-array beam --
in the form of 200 Nyquist-sampled, dual-polarization subbands of
195\,kHz each -- was processed using GPU-accelerated software to
perform coherent dedispersion and channelization with \textsc{cdmt}
\citep{bph17} at steps of 1\,pc\,cm$^{-3}$ between dispersion measures
(DMs) of 0.5 and 79.5\,pc\,cm$^{-3}$. The resulting coherent
filterbanks, sampled at 81.92\,$\upmu$s and 48.83\,kHz in time and
frequency, were dedispersed incoherently around the coherently
dedispersed DM trial ($\Delta\mathrm{DM}=-0.5$ to
$0.5$\,pc\,cm$^{-3}$) at steps of 0.002\,pc\,cm$^{-3}$ using the
\textsc{Dedisp} library \citep{bbbf12}. The dedispersed timeseries
were searched for periodic signals using frequency domain acceleration
searching (tools from \textsc{PRESTO}; \citealt{ran01,rem02}). A
detailed description of the processing steps is given in
\citet{bph17}.

PSR\,J0952$-$0607 was discovered blindly at high significance in 4 of
the 7 tied-array beams at a spin frequency of 707\,Hz and a dispersion
measure (DM) of 22.41\,pc\,cm$^{-3}$. The pulsar was found at an
acceleration of 1.3\,m\,s$^{-2}$, indicating that it is part of a
binary system. As the cumulative pulse profile of PSR\,J0952$-$0607 is
double peaked, with components separated by approximately
110\degr\ (see Fig.\,\ref{fig:profile}), we verified that the 707\,Hz
spin frequency is the fundamental (i.e.\ the neutron star's true
rotation rate) by folding the dedispersed timeseries of the discovery
observation at several harmonically related spin frequencies. In all
cases the resulting profiles were the sum of copies of the 707\,Hz
profile and were of lower signal-to-noise ratio than the 707\,Hz
profile.

Follow-up observations (10\,min integration times) were obtained with
the HBAs from 23 LOFAR core stations (longest baseline of 3.5\,km)
using 7 tied-array beams with 39\,MHz of bandwidth centered at
135\,MHz on 2017 January 4 (initial follow-up gridding observation),
and a single beam with the full HBA band (78\,MHz at 149\,MHz) for all
subsequent observations. These observations allowed us to refine the
position of the pulsar and start the timing program. A 3\,hr HBA
integration was obtained on 2017 January 28/29 to constrain the
orbital parameters. To determine the radio spectrum of
PSR\,J0952$-$0607, we obtained a 2\,hr integration with the LOFAR
low-band antennas (LBAs) between $30 - 90$\,MHz on 2017 February 5/6
and a 47\,min observation at 350\,MHz (100\,MHz bandwidth) on 2017
March 1 with GUPPI \citep{drd+08} at the Green Bank Telescope
(GBT). The pulsar was not detected in the LOFAR LBA observation, but
easily seen in the GBT 350\,MHz observation.

The complex voltage data of the discovery and follow-up LOFAR HBA
observations were coherently dedispersed and folded with
\textsc{dspsr} \citep{sb10} and analysed using \textsc{psrchive}
\citep{hsm04} tools. Pulse profiles for 2\,min sub-integrations were
referenced against an analytical pulse profile template to obtain
time-of-arrival (TOA) measurements. A phase-connected timing solution,
accounting for every rotation of the pulsar, was determined from these
TOAs using \textsc{tempo2} \citep{hem06,ehm06}.

\begin{deluxetable}{lc}
  \centering \tabletypesize{\footnotesize} \tablecolumns{2}
  \tablewidth{0pc} \tablecaption{Parameters for
    PSR\,J0952$-$0607.\label{tab:ephemeris}}
  \tablehead{\colhead{Parameters} & \colhead{Value}} \startdata
  \cutinhead{Timing Parameters} R.A., $\alpha_\mathrm{J2000}$ &
  $09^\mathrm{h}52^\mathrm{m}08\fs319(3)$ \\ Decl.,
  $\delta_\mathrm{J2000}$ & $-06\degr07\arcmin23\farcs49(5)$ \\ Spin
  frequency, $\nu$ (s$^{-1}$) & $707.314434911(16)$ \\ Spin frequency
  derivative, $\dot{\nu}$ (s$^{-2}$) & $>-3.3\times10^{-15}$ \\ Epoch
  of timing solution (MJD) & $57800$ \\ Dispersion measure, DM
  (pc\,cm$^{-3}$) & $22.41149(10)$ \\ Binary model & ELL1\\ Orbital
  period, $P_\mathrm{b}$ (d) & $0.267461038(12)$ \\ Projected
  semi-major axis, $x$ (s) & $0.0626694(14)$ \\ Time of ascending node
  passage, $T_\mathrm{asc}$ (MJD) & $57799.9119800(8)$ \\ Solar system
  ephemeris model & DE421 \\ Clock correction procedure & TT(BIPM2011)
  \\ Time Units & TCB\\ Timing Span (MJD) & 57747.1--57851.9\\ Number
  of TOAs & $164$ \\ Weighted rms post-fit residual ($\upmu$s) &
  5.6\\ Reduced $\chi^2$ value & 1.11\\ \enddata \tablecomments{The
    astrometric parameters ($\alpha_\mathrm{J2000}$ and
    $\delta_\mathrm{J2000}$) are kept fixed at the position of the
    optical counterpart. The eccentricity is kept fixed at $e=0$,
    implicitly assuming that the orbit is circular. For the ELL1
    binary model \citep{lcw+01}, this means $\kappa=e\sin\omega=0$ and
    $\eta=e\cos\omega=0$.}
\end{deluxetable}

\subsection{Optical}\label{ssec:optical}
We observed the field of PSR\,J0952$-$0607 using the Wide Field Camera
(WFC) on the 2.54\,m \textit{Isaac Newton Telescope} at the Roque de
Los Muchachos on La Palma. A dithered set of 240 2-min exposures with
a Sloan $r^\prime$ filter were obtained on 2017 January 17 and 18
under good conditions with $1\arcsec$ seeing. The WFC consists of four
$4\mathrm{k}\times2\mathrm{k}$ pixel CCDs, sampled at
$0\farcs33$\,pix$^{-1}$. In the following we use data from the center
chip, which contains the location of PSR\,J0952$-$0607. All images
were bias-subtracted and flat-fielded using dome flats and
subsequently registered using integer pixel offsets. To improve the
signal-to-noise ratio, we co-added between 5 and 20 images that were
consecutive in time.

We determined instrumental magnitudes through point-spread-function
(PSF) fitting using \textsc{DAOphot\,II} \citep{ste87} and calibrated
against $r^\prime$-band photometry from Pan-STARRS\,1 DR1
\citep{cmm+16,msf+16}. Astrometric positions from the GAIA DR1 catalog
\citep{bvp+16} were used for the astrometric calibration. A total of
58 GAIA stars overlapped with an $11\arcmin\times11\arcmin$ subsection
of a co-added image of 10 time consecutive 2\,min integrations. To
correct for the considerable distortion in the WFC camera, cubic
polynomials were used to relate pixel positions to right ascension and
declination. After iteratively removing two outliers, the astrometric
calibration yielded rms residuals of $0\farcs019$ in right ascension
and $0\farcs015$ in declination.

\begin{figure}
  \includegraphics[width=\columnwidth]{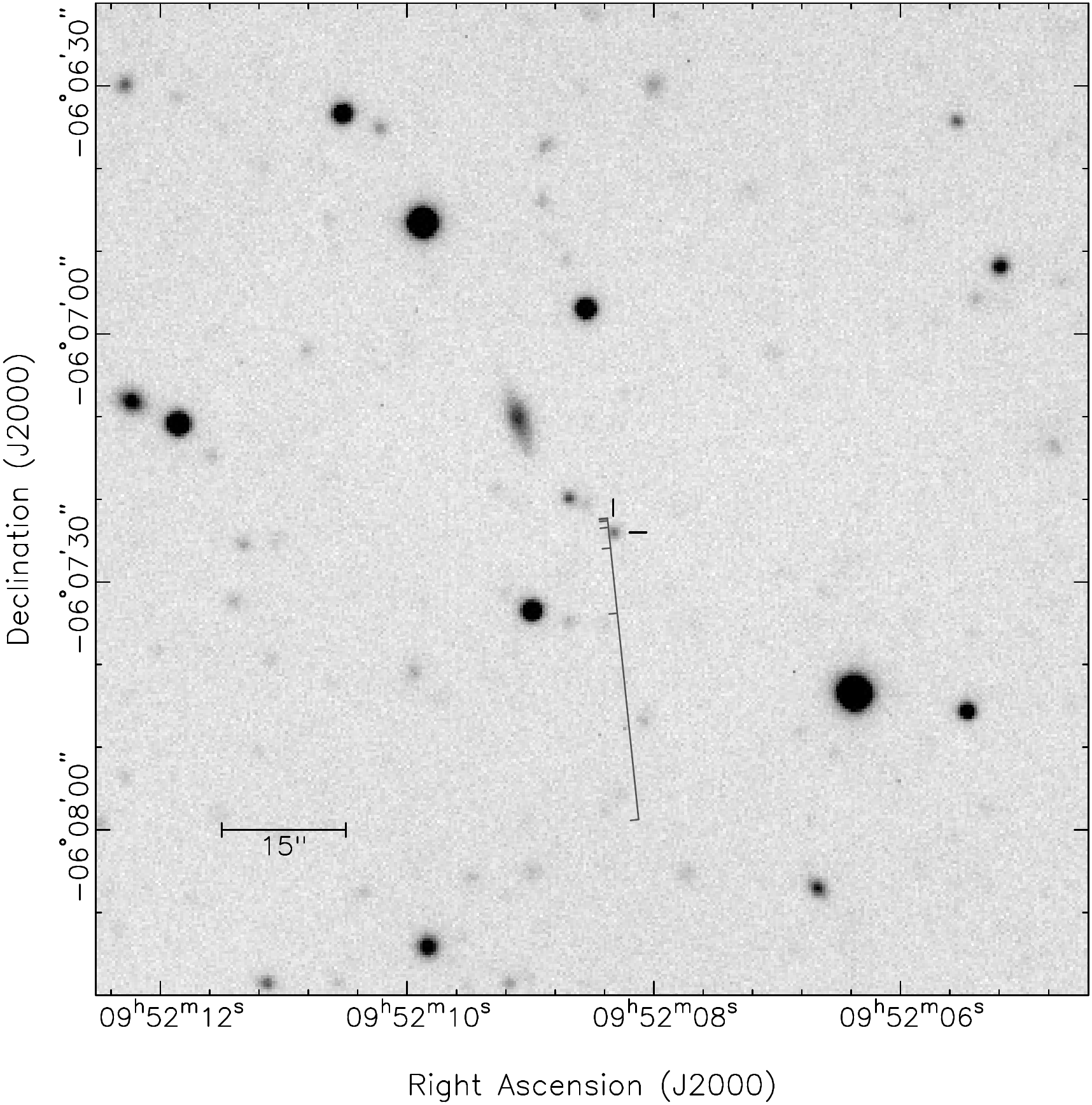}
  \caption{A $2\arcmin\times2\arcmin$ subsection of the
    $r^\prime$-band image consisting of 109 2\,min exposures between
    orbital phases $0.5<\phi<1.0$. The counterpart to
    PSR\,J0952$-$0607 is denoted by tickmarks ($2\arcsec$ in
    length). The diagonal line traces the position of timing
    ephemerides with spin frequency derivatives $\dot{\nu}$ between
    $-10^{-12}$ (bottom) to $-10^{-15}$\,s$^{-2}$ (top) in steps of
    0.5\,dex.}
  \label{fig:chart}
\end{figure}

\subsection{X-ray}\label{ssec:xray}
We obtained a 4.6\,ks \textit{Swift}/XRT observation of
PSR\,J0952$-$0607 on 2017\,March\,14 in photon-counting mode. The
\textsc{HEASOFT} tools were used for standard calibration and
extraction of events from a circular region with a radius of
71\arcsec\ and an annulus with inner and outer radii of 71 and
142\arcsec, centered on the position of the optical counterpart (see
below). Using standard response and exposure map calibration files, we
find that the count rate at the position of PSR\,J0952$-$0607 is
consistent with background noise, and that the X-ray counterpart to
PSR\,J0952$-$0607 is not detected.

\section{Results}\label{sec:results}
\begin{figure}
  \includegraphics[width=\columnwidth]{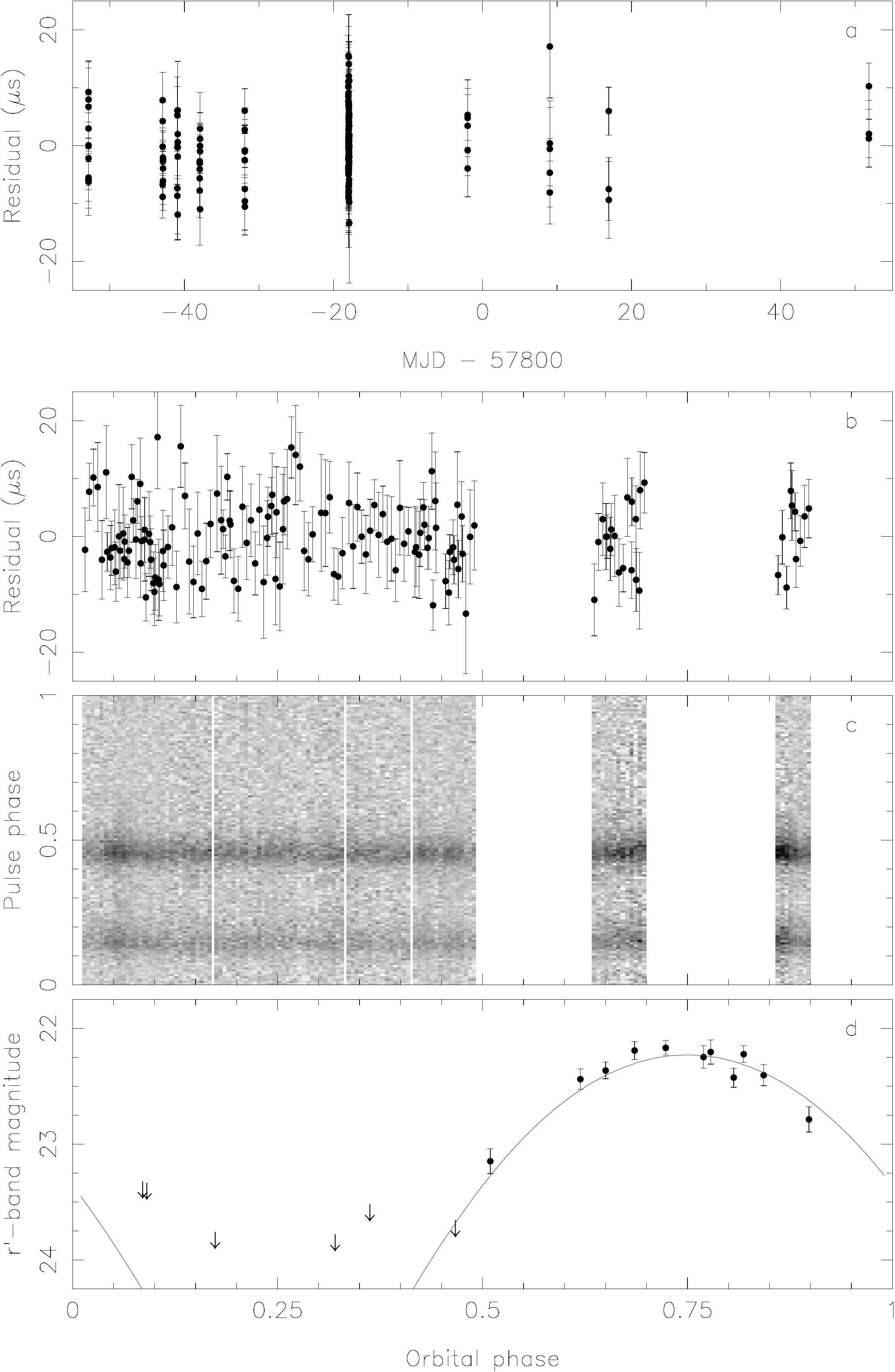}
  \caption{Timing residuals as a function of time and orbital phase
    are shown in panels a and b. Folded pulse profiles of
    PSR\,J0952$-$0607 are shown as a function of orbital phase for the
    orbital phases covered by our observations (panel c). Eclipses of
    the radio signal in black widow systems occur around orbital phase
    $\phi=0.25$ but are not obvious in PSR\,J0952$-$0607. Sloan
    $r^\prime$-band light-curve of the binary companion of
    PSR\,J0952$-$0607 (panel d). The \textsc{Icarus} model fit to
    the light curve is shown with the solid line.}
  \label{fig:lightcurve}
\end{figure}

The phase-connected timing solution models the rotation and orbit of
PSR\,J0952$-$0607. As the timing solution has a time baseline of
approximately a third of a year, the spin parameters ($\nu$ and
$\dot{\nu}$) are degenerate with the astrometric parameters
($\alpha_\mathrm{J2000}$, $\delta_\mathrm{J2000}$). Assuming values
for the spin frequency derivative $\dot{\nu}$ between $-10^{-12}$ to
$-10^{-15}$\,s$^{-2}$, the fitted position of the pulsar traces a line
on the sky as indicated in Fig.\,\ref{fig:chart}. Along this line
the optical images show a strongly variable object, located about
$1\arcmin$ from the LOFAR-gridded position of PSR\,J0952$-$0607. The
object varied by at least 1.5\,mags in the co-added $r^\prime$ images,
being below the detection threshold in approximately half of them. The
$r^\prime$-band magnitudes are modulated at the orbital period of
PSR\,J0952$-$0607 (Fig.\,\ref{fig:lightcurve}), confirming that the
object is the binary companion of the pulsar.

The binary companion to PSR\,J0952$-$0607 is located at
$\alpha_\mathrm{J2000}=09^\mathrm{h}52^\mathrm{m}08\fs319(3)$ and
$\delta_\mathrm{J2000}=-06\degr07\arcmin23\farcs49(5)$. The positional
uncertainty quoted is the quadratic sum of the uncertainty in the
astrometric calibration and the positional uncertainty of the
companion on the co-added image (of order $0\farcs05$). To estimate
the impact of the uncertainty in the optical position of the binary
companion on the parameters of the timing solution, we performed a
Monte Carlo simulation, drawing positions ($\alpha_\mathrm{J2000}$,
$\delta_\mathrm{J2000}$) from normal distributions with appropriate
means and widths. For each of these positions, the remaining
parameters in the timing solution were fitted to generate
distributions from which the parameters and their uncertainties were
determined. These values are listed in
Table\,\ref{tab:ephemeris}. Taking into account the positional
uncertainties yields a 3-$\sigma$ limit on the spin frequency derivative
$\dot{\nu}>-3.3\times10^{-15}$\,s$^{-2}$, corresponding to a spin
period derivative of $\dot{P}<1.1\times10^{-20}$\,s\,s$^{-1}$, placing
it in the lower half of the MSP $\dot{P}$ distribution. Because of the
short spin period, the surface magnetic field
$B\propto\sqrt{P\dot{P}}$ is low, $B<1.3\times10^8$\,G. The limit on
the spindown luminosity ($\dot{E}\propto\dot{P}P^{-3}$) is
$\dot{E}<1.6\times10^{35}$\,erg\,s$^{-1}$.
%We note that these values
%could be significantly lower should PSR\,J0952$-$0607 have a
%significant proper motion, as the observed $\dot{P}$ would have a
%contribution due to the Shklovskii effect \citep{shk70}.

Models for the electron distribution in our Galaxy along the
line-of-sight to PSR\,J0952$-$0607 ($l=243\fdg65$, $b=35\fdg38$)
constrain the distance through the observed dispersion measure
(DM). The NE2001 model by \citet{cl02} predicts a distance of
$d=0.97$\,kpc, while the YMW16 model \citep{ymw17} places it
significantly farther away at $d=1.74$\,kpc. At either distance, the
Galactic extinction model by \citet{gsf+14,gsf+15} estimates the same
reddening of $E_{B-V}=0.061$, leading to an extinction of
$A_{r^\prime}=0.14$ \citep{sf11}.

The orbital parameters from the timing solution
(Table\,\ref{tab:ephemeris}) show that PSR\,J0952$-$0607 is in a
6.42-hr binary around an ultra low-mass companion. The mass function
is $3.69\times10^{-6}$\,M$_\odot$, setting the minimum companion mass
at 0.019\,M$_\odot$ for a 1.4\,M$_\odot$ pulsar. These properties are
consistent with PSR\,J0952$-$0607 being a black widow system, where
matter from the companion is ablated by the energetic pulsar wind
(e.g.\ \citealt{fst88}, see \citealt{rob13} for a review).

The black widow nature of PSR\,J0952$-$0607 is confirmed by the
sinusoidal optical light curve, which is consistent with irradiation
of the companion hemisphere facing the pulsar. We use the
\textsc{Icarus} software \citep{brkc12} to model the $r^\prime$-band
light curve. As the absence of color information precludes a full
parameter fit, we make the following assumptions to estimate system
parameters. We assume that the companion is co-rotating and that its
base temperature (that of the unirradiated hemisphere) is 2500\,K --
in line with what is seen in other black widows (see,
e.g.\ \citealt{kbk11,skbk01,rgfz16}) -- and hence contributes
minimally to the flux of the irradiated hemisphere. The reddening is
kept at $E_{B-V}=0.061$ (see above), and the pulsar mass is fixed at
1.4\,M$_\odot$. Finally, we require the dayside temperature to be such
that the implied irradiation represents of order 10\% of the spin down
luminosity for an assumed credible range from $5\times10^{34}$ to
$1.6\times10^{35}$\,erg\,s$^{-1}$ \citep{bkr+13}.

We find that the observed light curve is mostly inconsistent with
models that assume a filling factor of unity (i.e.\ where the
companion fills the Roche lobe), as it leaves the orbital inclination
largely unconstrained, though leaning towards edge on ($i=90\degr$),
with tightly correlated dayside temperatures and distances in the
range of 4500 to 5800\,K and 2.3 to 5.3\,kpc, respectively. The
goodness of fit improves significantly for models with an assumed
filling factor of 0.5, providing a well defined inclination of
$i\sim40\degr$. The correlation between dayside temperature and
distance is weaker, yielding similar dayside temperatures and slightly
smaller distances (1.7 to 3.8\,kpc). Models with the filling factor as
a free parameter prefer slightly smaller filling factors while the
overall goodness of the fit is not significantly increased. These
models provide similar constraints on the inclination, dayside
temperature and distance, with even weaker correlations due to the
extra free parameter.

We conclude that PSR\,J0952$-$0607 is almost certainly not close to
Roche-lobe filling, with an orbital inclination in the intermediate
range. The predicted model distances are at the high end of those
estimated from the DM, which may suggest Roche lobe filling factors of
0.5 or less, or lower dayside temperatures, indicating spindown
luminosities of a few $10^{34}$\,erg\,s$^{-1}$ or that the conversion
of spindown into heating is less than 10\% efficient. Upcoming
multi-color photometry will be able to confirm these values.

The low Roche lobe filling factor and low inclination are consistent
with the absence of eclipses of the radio signal in the observations
obtained so far. Eclipses are seen in the majority of black widow
systems, and eclipses are generally most pronounced at low observing
frequencies (e.g.\ \citealt{sbl+96,asr+09}). Besides eclipses, ionized
matter passing through the line-of-sight leads to increases in the DM
near orbital phase $\phi=0.25$, resulting in delays in the TOAs. The
TOA residuals in Fig.\,\ref{fig:lightcurve} do show systematic delays
at $\phi=0.26$ to 0.28, possibly hinting at the presence of ionized
material in the line-of-sight. The ongoing LOFAR timing observations
will constrain whether material is ablated from the companion of
PSR\,J0952$-$0607.

We use the \citet{ham06} model for the LOFAR HBA beam together with
the radiometer equation and the method as detailed in \citet{kvh+16},
to obtain flux density measurements for PSR\,J0952$-$0607 over the HBA
band between 110 to 188\,MHz. Measurements from 5 different
observations, totalling 1\,hr of integration time, were averaged. The
flux density at 350\,MHz was measured from the single GBT observation,
using the radiometer equation with gain, system temperature and
bandwidth values from \citet{slr+14}. We obtain $S_\mathrm{mean}=45$,
32, 21, 9 and 1.5\,mJy at frequencies of 119.6, 139.2, 158.7, 178.2
and 350\,MHz. Following \citet{bkk+16}, we conservatively estimate
50\% uncertainties in the flux density measurements. Modelling the
spectrum with a power law $S_\nu\propto\nu^\alpha$ over frequency
$\nu$ yields a spectral index of $\alpha=-3.3\pm0.3$ and a flux
density at 150\,MHz of $S_{150}=21\pm2$\,mJy. We note that
PSR\,J0952$-$0607 is not detected in the 150\,MHz TGSS-ADR source
catalog \citep{ijmf17}, listing sources brigther than $7\sigma$
significance with a median noise of 3.5\,mJy\,beam$^{-1}$. At the
location of PSR\,J0952$-$0607 the flux density in the TGSS-ADR images
is 12\,mJy\,beam$^{-1}$, suggesting a $3\sigma$ detection, within
$2\sigma$ from the LOFAR flux density. Even if the LOFAR fluxes are
overestimated by a factor of two, as suggested by \citet{fjmi16}, the
spectral index remains as steep as $\alpha=-2.6\pm0.4$.

Figure\,\ref{fig:profile} shows the cumulative pulse profile of
PSR\,J0952$-$0607 at different frequencies, revealing significant
evolution with observing frequency. The pulse profile evolves from two
double peaked components at an observing frequency of 350\,MHz to two
scatter-broadened components at LOFAR frequencies. To estimate the
scattering time scale, we approximate scatter broadening as a
convolution with a truncated exponential, appropriate for a thin
scattering screen \citep{wil72}, and assume that scatter broadening
can be neglected in the 350\,MHz pulse profile such that it can be
treated as the intrinsic profile. The four components are modeled with
von Mises functions and the amplitude of the four components was
allowed to vary with frequency, while positions and widths are kept
fixed. Using this model we find scattering times of 47 and
113\,$\upmu$s at observing frequencies of 168 and 129\,MHz,
respectively. The first two components of the pulse profile, measured
against the highest component, increase by a factor 1.9 and 2.0 from
350 to 129\,MHz, while the third component decreases by a factor 3.3.

These scattering time scales indicate that the scatter broadening
exceeds the pulse period for observing frequencies below 70\,MHz,
assuming scattering scales as $\tau\propto\nu^{-4}$. As the LBA
sensitivity peaks near 60\,MHz, where the pulsed signal will be
scattered out completely, it is not surprising that PSR\,J0952$-$0607
is not detected with the LOFAR LBA.

The \textit{Swift}/XRT X-ray non-detection of PSR\,J0952$-$0607
translates to a $3\sigma$ flux limit in the $0.3 - 10$\,keV band of
$f_\mathrm{X}<1.1\times10^{-13}$\,erg\,s$^{-1}$\,cm$^{-2}$ for
absorbed blackbody ($T_\mathrm{eff}=0.23$\,keV) and powerlaw
($\Gamma=2$) spectra. Here, we assumed
$N_\mathrm{H}=4\times10^{20}$\,cm$^{-2}$ estimated from the optical
reddening through the relation by \citet{go09}. The resulting
$3\sigma$ X-ray luminosity limits ($0.3-10$\,keV) are
$L_\mathrm{X}<1.1\times10^{31}$\,erg\,s$^{-1}$ at a distance of
0.97\,kpc and $L_\mathrm{X}<3.6\times10^{31}$\,erg\,s$^{-1}$ at
1.74\,kpc. These limits are consistent with the observed relation
between the X-ray luminosity and spindown luminosity of radio MSPs
\citep{pccm02}.

\section{Discussion and conclusions}\label{sec:discussion}
PSR\,J0952$-$0607 has a spin frequency $\nu = 707$\,Hz. This makes it
the fastest-spinning neutron star known in the Galactic field (outside
of a globular cluster), surpassing the 35\,year record set by the
first MSP to be discovered, PSR\,B1937+21, which spins at 642\,Hz
\citep{bkh+82}. Only PSR\,J1748$-$2446ad, located in the globular
cluster Terzan~5, spins faster at 716\,Hz \citep{hrs+06}.  Of the 213
known Galactic field MSPs with $P<30$\,ms ($\nu > 33$\,Hz) from the
pulsar
catalog\footnote{\url{http://www.atnf.csiro.au/people/pulsar/psrcat}}
  \citep{mhth05}, only 13 have $P < 2$\,ms ($\nu > 500$\,Hz).  A
  further 3 MSPs in globular clusters also satisfy this condition,
  including PSR\,J1748$-$2446ad. Of the accreting millisecond X-ray
  pulsars, only 3 out of 15 have spin frequencies above 500\,Hz
  \citep{pw12}.

For PSR\,J1748$-$2446ad it is not possible to determine the intrinsic
$\dot{\nu}$ of the neutron star because the observed change in
spin-rate with time is dominated by acceleration in the gravitional
potential of Terzan~5 \citep{prf+17}.  Conversely, it will be possible
to measure PSR\,J0952$-$0607's intrinsic spin-down rate and the
inferred surface magnetic field, once a full timing solution is
available. These measurements will shed light on the question of
whether the fastest-spinning radio MSPs also typically have the lowest
magnetic fields.  The current limit of $B<1.3\times10^8$\,G already
qualifies PSR\,J0952$-$0607 as one of the most weakly magnetized
pulsars known.

Including PSR\,J0952$-$0607, there are 14 Galactic radio MSPs with
$\nu>500$\,Hz, of which: 5 are in black widow systems, 4 are isolated
pulsars, 3 are in redback systems, and 2 have white dwarf companions.
The abundance of black widow and redback systems amongst the
fastest-spinning MSPs may hint at an evolutionary origin, possibly
related to the accretion process and the amount of accreted matter
\citep{hess08,ptrt14}\footnote{Furthermore, it has been suggested that
  the isolated MSPs represent the outcomes of extreme black widow
  systems, in which the pulsar wind has successfully evaporated the
  companion star entirely \citep{fst88}.}. We note that radial velocity
measurements and light curve modeling of the black widow companions to
PSR\,B1957+20 ($\nu=622$\,Hz) and PSR\,J1301+0833 ($\nu=543$\,Hz),
indicate that the MSPs are heavy, with masses of
$2.40\pm0.12$\,M$_\odot$ \citep{kbk11} and
$1.74_{-0.17}^{+0.20}$\,M$_\odot$ \citep{rgfz16}, respectively. Future
photometric and spectroscopic observations of the companion of
PSR\,J0952$-$0607 could test this hypothesis.

Besides the short spin period, PSR\,J0952$-$0607 is remarkable due to
its steep radio spectrum. At $\alpha\sim-3$, its spectral index is
amongst the steepest known compared to recent studies by
\citet{kvl+15}, \citet{kvh+16} and \citet{fjmi16}. Our LOFAR and GBT
observations should be robust against scintillation, as the bandwidths
and integration times used substantially exceed the scintillation
bandwidth and timescale towards PSR\,J0952$-$0607 \citep{cl02}.

The steep spectrum of PSR\,J0952$-$0607 adds to the emergent picture
where the fastest-spinning MSPs tend to have the steepest spectra, but
also that the steepest spectra MSPs tend to be detected by
\textit{Fermi} in $\gamma$-rays \citep{kvl+15,fjmi16}. Given that the
fastest-spinning radio MSPs tend to have aligned radio and
$\gamma$-ray profiles \citep{egc+13,jvh+14}, it is suggestive that
these tendencies are pointing to a commonality in the radio and
$\gamma$-ray emission mechanism, where the fast spin frequency leads
to emission of $\gamma$-rays co-located with steep spectrum radio
emission. Continued radio timing of PSR\,J0952$-$0607 will allow us to
fold the \textit{Fermi} $\gamma$-ray photons, and test the alignment
between the radio and $\gamma$-ray profiles.

The discovery of PSR\,J0952$-$0607 and the previous LOFAR discovery of
PSR\,J1552+5437 (\citealt{bph17,pbh+17}) demonstrate the potential of
MSP searches at unconventionally low radio observing frequencies.
Such searches are more sensitive to the population of
ultra-steep-spectrum radio MSPs ($\alpha < -2.5$), and can explore
whether even faster-spinning, potential sub-millisecond pulsars, can
be formed in nature but have previously been missed at higher
observing frequencies.  Unfortunately, low-frequency searches still
suffer from limitations due to IISM scattering and eclipses. However
recent results have demonstrated the power of selecting such sources
in low-frequency radio interferometric imaging surveys, where such
effects do not affect the detectability of the source, and then
following up with deep time domain searches \citep{fjmi16}. Both
imaging-led and direct time-domain search approaches should continue
to be exploited.

While there is currently no clear theoretical expectation that radio
spectral index should depend on spin rate, we note that the shrinking
size of the light cylinder $r_c = 48 (P/1~{\rm ms})$\,km could
plausibly play a role in spectral steepening because the
magnetospheric size becomes comparable to, or smaller compared to the
typical emission height.  Furthermore, it is interesting to consider
whether the radio emission from the fastest-spinning MSPs is dominated
by giant pulse emission from a region co-located with the high-energy
$\gamma$-ray emission.

\acknowledgments We thank LOFAR Science Operations and Support for
their help in scheduling and effectuating these observations. We also
thank Caroline D'Angelo, Gemma Janssen and Alessandro Patruno for
useful discussions. CGB and JWTH acknowledge support from the European
Research Council (ERC) under the European Union's Seventh Framework
Programme (FP/2007-2013) / ERC Grant Agreement nr.\ 337062 (DRAGNET;
PI Hessels). RPB had received funding from the ERC under the European
Union's Horizon 2020 research and innovation programme (grant
agreement nr.\ 715051; Spiders). This paper is based on data obtained
with the International LOFAR Telescope (ILT) under projects LC7\_002,
DDT7\_002 and LT5\_003. LOFAR is the Low Frequency Array designed and
constructed by ASTRON. It has facilities in several countries, that
are owned by various parties (each with their own funding sources),
and that are collectively operated by the ILT foundation under a joint
scientific policy. The Isaac Newton Telescope and its service
programme are operated on the island of La Palma by the Isaac Newton
Group of Telescopes in the Spanish Observatorio del Roque de los
Muchachos of the Instituto de Astrof\'isica de Canarias. The
\textit{Fermi} LAT Collaboration acknowledges generous ongoing support
from a number of agencies and institutes that have supported both the
development and the operation of the LAT as well as scientific data
analysis.  These include the National Aeronautics and Space
Administration and the Department of Energy in the United States, the
Commissariat \`a l'Energie Atomique and the Centre National de la
Recherche Scientifique / Institut National de Physique Nucl\'eaire et
de Physique des Particules in France, the Agenzia Spaziale Italiana
and the Istituto Nazionale di Fisica Nucleare in Italy, the Ministry
of Education, Culture, Sports, Science and Technology (MEXT), High
Energy Accelerator Research Organization (KEK) and Japan Aerospace
Exploration Agency (JAXA) in Japan, and the K.~A.~Wallenberg
Foundation, the Swedish Research Council and the Swedish National
Space Board in Sweden. Additional support for science analysis during
the operations phase is gratefully acknowledged from the Istituto
Nazionale di Astrofisica in Italy and the Centre National d'\'Etudes
Spatiales in France. This work was performed in part under DOE
Contract DE-AC02-76SF00515.

\facilities{LOFAR, ING:Newton, Swift, GBT, Fermi}
\software{cdmt, PRESTO, PSRCHIVE, Astropy, ESO-MIDAS, HEASOFT}

\bibliographystyle{aasjournal}
%\bibliography{references}

\end{document}